\documentclass[12pt,a4]{article}
\usepackage{epsfig}
\usepackage{setspace}
\newcommand{\beq}{\begin{equation}}
\newcommand{\eeq}{\end{equation}}
\newcommand{\bea}{\begin{eqnarray}}
\newcommand{\eea}{\end{eqnarray}}
\newcommand{\ba}{\begin{array}}
\newcommand{\ea}{\end{array}}

\title{Considerations for Anderson-Bridge Experiment}
\author{ P. Arun,  Kuldeep Kumar \& Mamta\\
{\sl Department of Physics and Electronics},\\
{\sl S.G.T.B. Khalsa College}\\
{\sl University of Delhi, Delhi 110 007, India.}}

\begin{document}
\maketitle
\begin{abstract}
Various issues concerned with the design and 
sensitivity of the Anderson bridge
are discussed. We analyse the circuit using the tools of Network
Analysis to determine the conditions under which the balance may be
obtained with greater sensitivity and discuss how to obtain these
under practical circumstances in an undergraduate laboratory.

\end{abstract}

\section{Introduction}

\begin{figure}[t!!]
\begin{center}
\includegraphics[width=3.5in,angle=0]{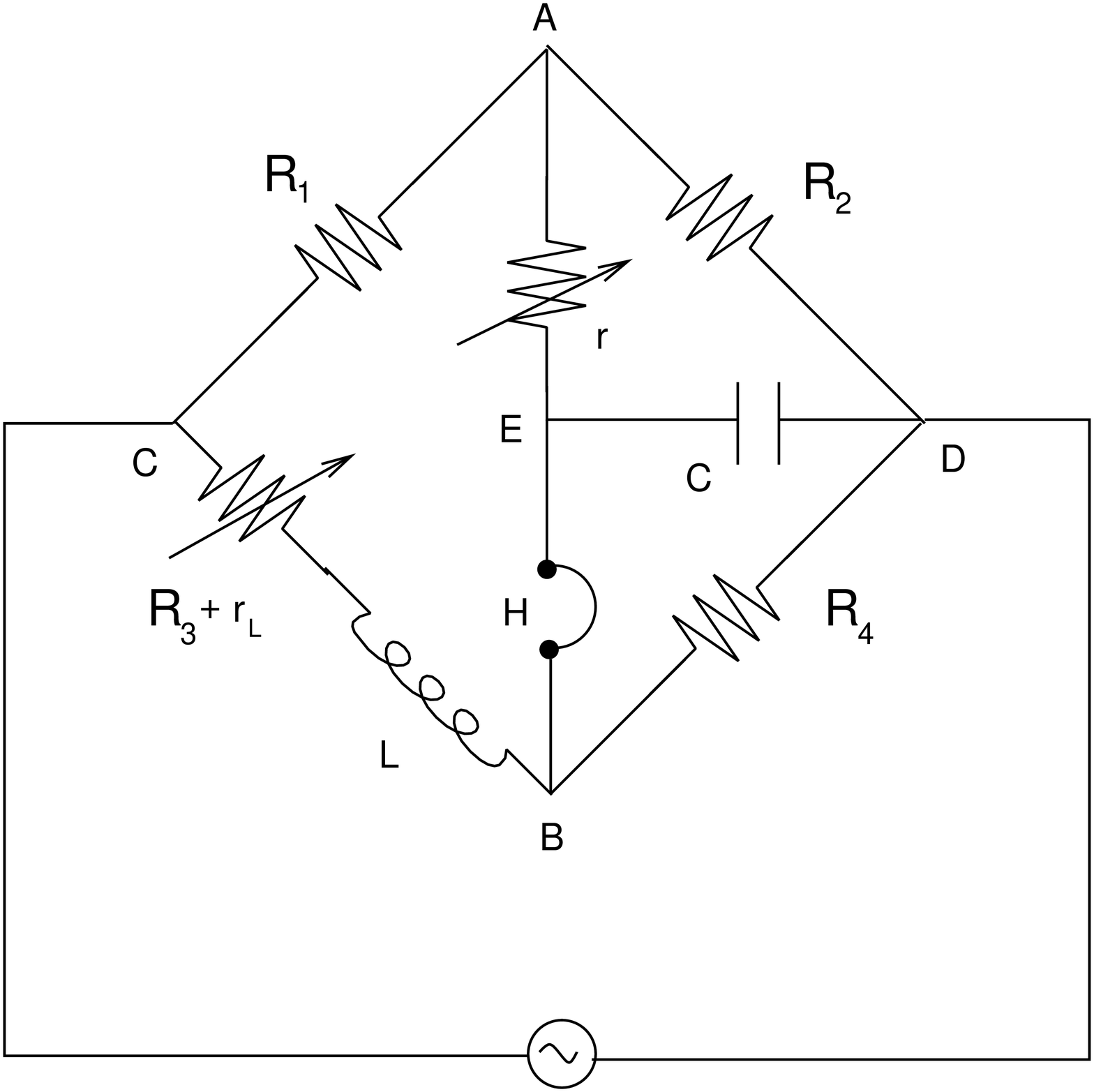}
\caption{The circuit schematics for Anderson Bridge. The resistance ${\rm
r_L}$ is the resistance of the coil whose self-inductance is being
determined.}  
\label{fig_and} 
\end{center}
\end{figure}

AC bridges are often used to measure the value of an unknown impedence
for example self/mutual inductance of inductors or capacitance of
capacitors accurately. A large number of AC bridges are available for the
accurate measurement of impedences. An Anderson Bridge is used to measure
the self inductance of a coil (Fig.~\ref{fig_and}). This is an old experiment 
and has been a part of the graduation curriculum since ages. In fact the
oldest publication on sensitivity of A.C. bridges is Rayleigh's
paper~\cite{rayleigh}. As it usually happens in such old subjects,
modern textbooks have diluted the attention paid to the details and
intricacies of the experiment. Most of the textbooks tend to merely
state the balance condition without discussing the design of the
experiment for greater sensitivity.

Over the years of instructing young graduate students in the lab, the
authors have observed unsatisfactory results reported by the students 
in terms of accuracy. One reason reported by the
students was an inability to get a mute on balance condition in the
head-phone. That is, the human perception of point of minima rendered 
results inaccurate. The fact that human ears perceive in decibels makes the
situation worse. A question that invariably arose was whether ``urbanisation
and sound pollution was effectively contributing to the inaccuracies
of this experiment (physiological constraints) or rather it was 
ignorance of the relevant physics contributing.

As a test experiment, a careful experimentalist from among the
under-graduate students was given basic instructions and a copy of
Yarwood's book~\cite{yarwood} for review before conducting the
Anderson Bridge experiment. Table 1, 2 and 3 list the results reported
by this student.

\begin{table}[h]
\begin{center}
\caption{\sl Table lists the experimentally determined value of
inductance (L) in bridge with coil L=207.4mH and DC balanced resistances
${\rm R_1=560\Omega}$, ${\rm R_2=552\Omega}$ and ${\rm
R_4=1490\Omega}$.}
\vskip 0.5cm
\begin{tabular}{|c|c| c| c| c| }
\hline
S.No. & C (${\rm \mu F}$) & r (${\rm \Omega}$) & ${\rm r_{av}}$ (${\rm
\Omega}$) & L (mH) \\ \hline
1. & 0.05 & 1000-1050 & 1025 & 195.55 \\ \hline
2. & 0.10 & 400-450 & 425 & 211.00  \\ \hline
3. & 0.11 & 390-410 & 350 & 217.25  \\ \hline
4. & 0.12 & 200-250 & 225 & 181.17 \\ \hline
5. & 0.15 & 170-200 & 185 & 208.45  \\ \hline
6. & 0.20 & 43-55 & 49 & 196.30  \\ \hline
\end{tabular}
\label {tab}
\end{center}
\vskip -0.75cm
\end{table}

\begin{table}[h]
\begin{center}
\caption{\sl Table lists the experimentally determined value of
inductance (L) in bridge with coil L=240.2mH and DC balanced resistances
${\rm R_1=560\Omega}$, ${\rm R_2=552\Omega}$ and ${\rm
R_4=1490\Omega}$.}
\vskip 0.5cm
\begin{tabular}{|c|c| c| c| c| }
\hline
S.No. & C (${\rm \mu F}$) & r (${\rm \Omega}$) & ${\rm r_{av}}$ (${\rm
\Omega}$) & L (mH) \\ \hline
1. & 0.05 & 1250-1310 & 1280 & 233.83 \\ \hline
2. & 0.10 & 485-515 & 500 & 233.52  \\ \hline
3. & 0.11 & 390-400 & 395 & 222.20  \\ \hline
4. & 0.15 & 300-310 & 305 & 209.99  \\ \hline
5. & 0.20 & 105-115 & 110 & 232.92  \\ \hline
\end{tabular}
\label {tab}
\end{center}
\vskip -0.75cm
\end{table}

\begin{table}[h!!]
\begin{center}
\caption{\sl Table lists the experimentally determined value of
inductance (L) in bridge with coil L=262.9mH and DC balanced resistances
${\rm R_1=560\Omega}$, ${\rm R_2=552\Omega}$ and ${\rm
R_4=1490\Omega}$.}
\vskip 0.5cm
\begin{tabular}{|c|c| c| c| c| }
\hline
S.No. & C (${\rm \mu F}$) & r (${\rm \Omega}$) & ${\rm r_{av}}$ (${\rm
\Omega}$) & L (mH) \\ \hline
1. & 0.05 & 1460-1510 & 1485 & 264.59 \\ \hline
2. & 0.10 & 650-700 & 675 & 286.05  \\ \hline
3. & 0.11 & 480-510 & 495 & 255.22  \\ \hline
4. & 0.12 & 390-410 & 400 & 244.21 \\ \hline
5. & 0.20 & 100-160 & 130 & 244.92  \\ \hline
\end{tabular}
\label {tab}
\end{center}
\vskip -0.75cm
\end{table}

As can be seen on examination of the results listed in the Tables, the
results are inaccurate and are scattered and deviated from the known
value. Also, notice that the student was not able to resolve the minimum
for a range of `r' values (of the order of ${\rm 50\Omega}$). The
results reported by the remaining students of the class by and large
suffered from more
inaccuracies. Under such circumstances, it warrants a more serious
analysis of the experiment and in process polish our understanding of
the same which with time seems to have lost relevance. Our analysis is
markedly different from that listed by Yarwood~\cite{yarwood} and
therewith cited Rayleigh's work~\cite{rayleigh}. However, it
utilizes the basic theorems taught in Network Analysis and if not
considered a serious contribution, it may be viewed as a
different preceptive and an attempt in highlighting the
utility of the various Network Theorem in a simple circuit.

Before dwelling on the mathematics and circuit design considerations, for 
completeness and rendering the article self sufficient for the reader, we
explain in short here, the Wheatstone Bridge and standard steps followed for 
DC and AC balancing of the bridges for the determination of the unknown
impedences. 

\begin{figure}[t!!]
\begin{center}
\includegraphics[width=3.5in,angle=0]{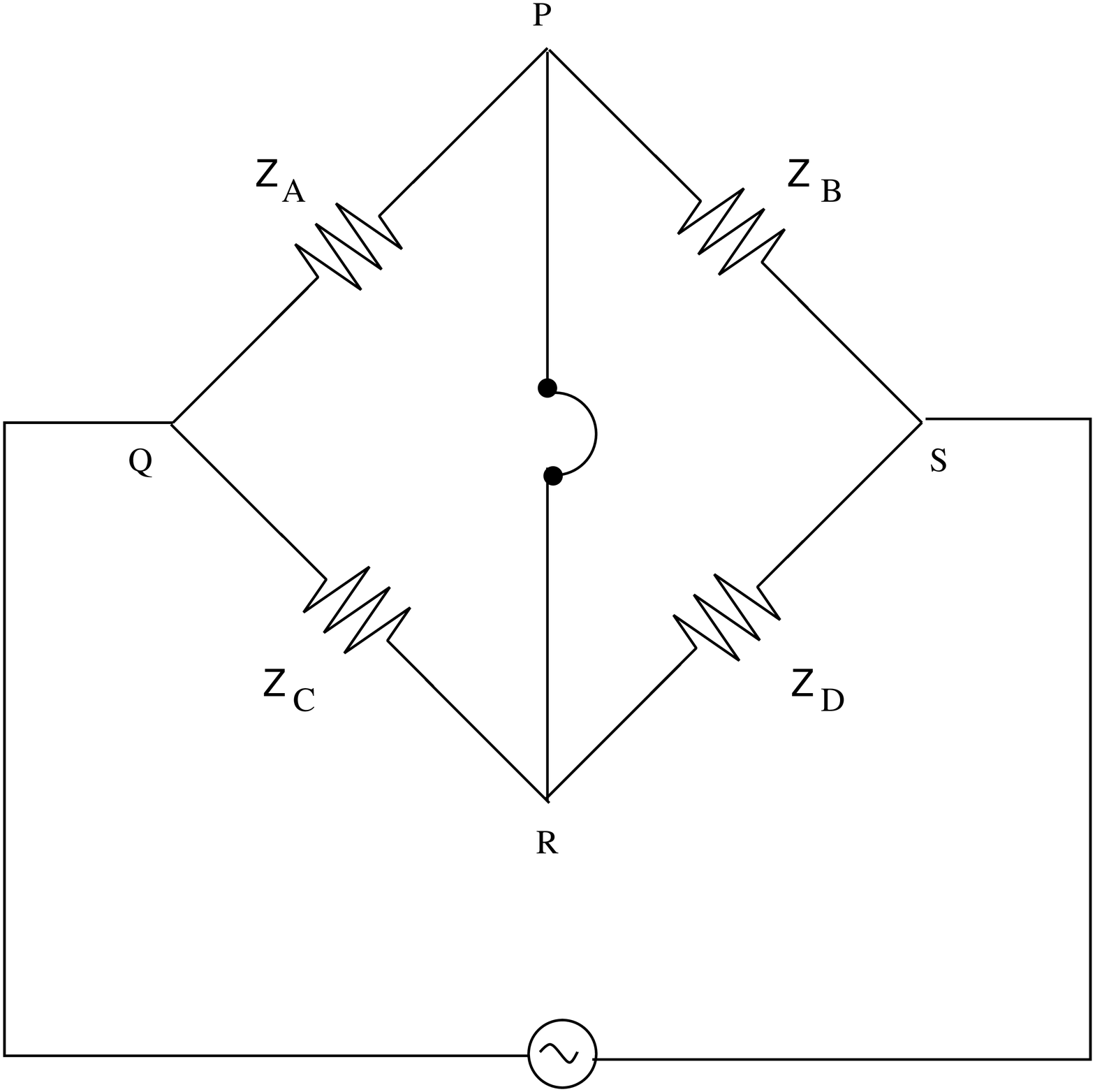}
\caption{A simple Wheatstone Bridge Arrangement.}
\label{fig_wh}   
\end{center}
\end{figure}

Most of the AC bridges are based
on a generalised Wheatstone Bridge circuit. As shown in Fig.~\ref{fig_wh}, the four arms of the D.C. Wheatstone Bridge are replaced
by impedences ($Z_A$, $Z_B$, $Z_C$ and $Z_D$), the battery by an
A.C. source and the D.C. galvanometer by an A.C. null detector
(usually a pair of headphones).
Using Kirchoff's Laws, it can be easily shown that the balance or null
condition (i.e. when no current flows through the detector or the
potential at the point P becomes equal to that at point R) is given
by
\beq
\frac{Z_A}{Z_B} = \frac{Z_C}{Z_D}. 
\label{wh_bal}
\eeq
Eqn~(\ref{wh_bal}) is a complex equation i.e. it represents two
real equations obtained by separately equating the real and imaginary
parts of the two sides. It follows from
the fact that both amplitude and the phase must be balanced. 
This implies that to reach the balance
condition two different adjustments must be made. That is, the DC and AC 
balance conditions have to be obtained one
by one. The DC balance is obtained using a D.C. source and moving coil 
galvanometer by adjusting one of the resistances. After which the battery and 
the galvanometer are replaced with 
an AC source and a headphone respectively, without changing the resistances
set eariler,  
the other variable impedence is varied to obtain minimum sound in the 
headphone.

\section{Circuit Designing}

Various combinations of the impedences can satisfy the balance
conditions of a bridge. For example, in a D.C. Wheatstone Bridge 
using either $R_A = R_B=R_C =R_D = 
10 \,\Omega $ or $R_A = R_B = 10 \,\Omega$ and
$R_C =R_D = 1000 \,\Omega $ would balance the
bridge as indicated by eqn.~(\ref{wh_bal}). However, the
bridge is not equally sensitive in both cases. Also interchanging the
positions of the source and detecting instrument do not alter the
balance conditions (Reciprocity Theorem) but again the bridge may not
be equally sensitive in the two cases \cite{yarwood}. The bridge will be 
more
`sensitive' if, in a balanced bridge, changing the impedence in any one
of the arms of the bridge from the value $Z$ required by the balance
conditions to value $Z+\delta Z$ (making the bridge off balance)
results in a greater current through the detecting instrument. As explained
by Yarwood~\cite{yarwood}, a D.C. bridge is more sensitive
if, {\sl whichever has the greater resistance, galvanometer or battery, is
put across the junction of the two lower resistances to the junction
of the two higher resistances. Also the unknown resistance should be
connected in the Wheatstone bridge between a small ratio arm
resistance and a large known variable resistance}. However, it is
common practice to make all the four arms of Wheatstone Bridge equal 
or make them of the same order for the optimum
sensitivity. For A.C. bridges also, the same procedure is followed. Below we
discuss designing an Anderson Bridge of decent sensitivity.

Most text books derive the DC and AC balance conditions of the Anderson Bridge
using Kirchoff's current and voltage laws. As stated earlier, we plan to use
various theorems taught in a course of {\sl Network Analysis} to do the
same. Hence, to analyse the bridge, we convert the ${\rm \pi}$ network 
consisting of $r$, $ R_2$, and $ X_c$ and replace it with an 
equivalent `T' network, whose three impedances would be given as
\cite{edmin}
\begin{eqnarray}
Z_1 &=& { rR_2 \over r+R_2+X_c}\label{z1}\\
Z_2 &=& { R_2X_c \over r+R_2+X_c}\label{z2}\\
Z_3 &=& { rX_c \over r+R_2+X_c}\label{z3}
\end{eqnarray}
\begin{figure}[t!!]
\begin{center}
\includegraphics[width=2.2in,angle=0]{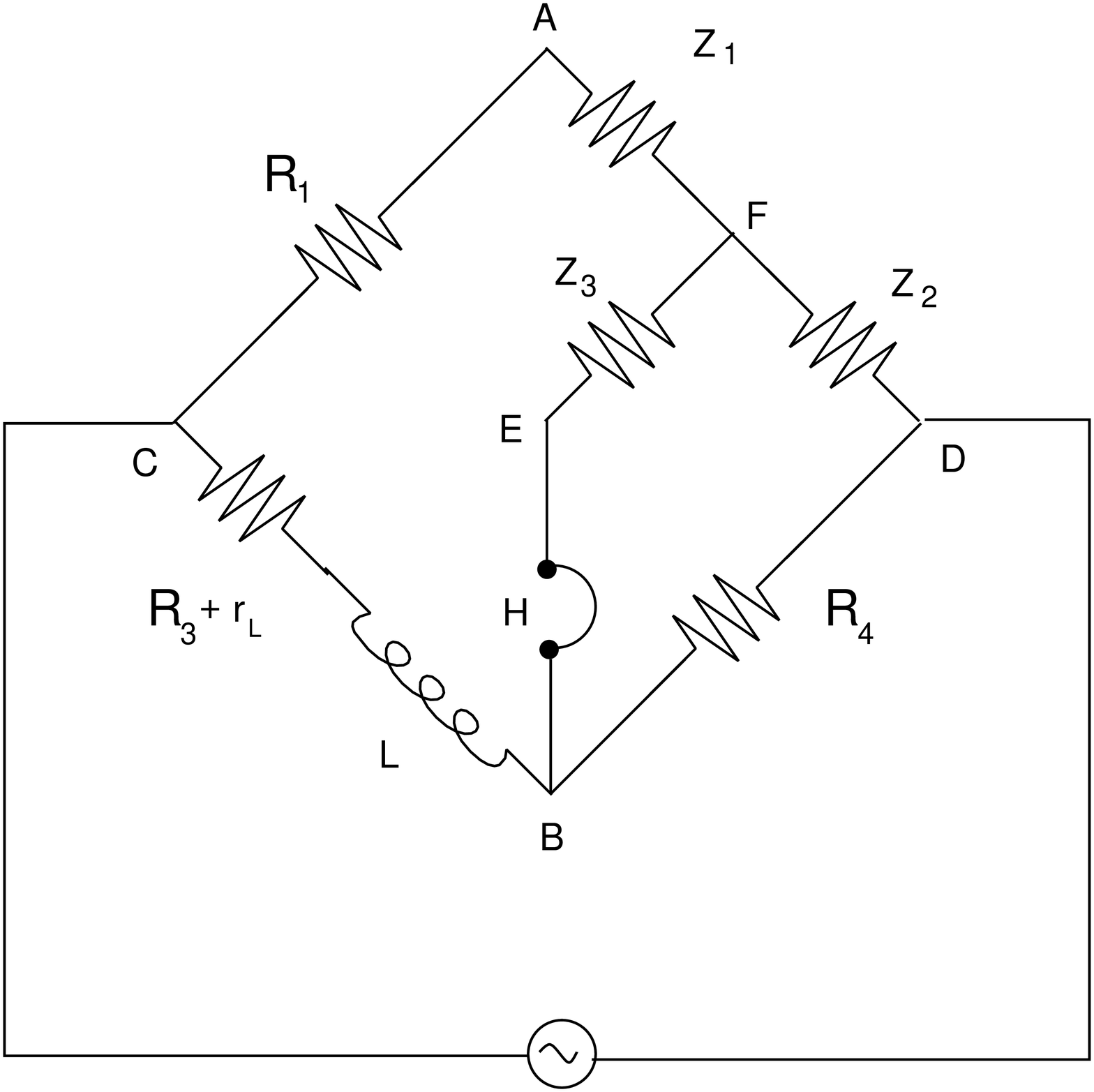}
\hfil
\includegraphics[width=2.2in,angle=0]{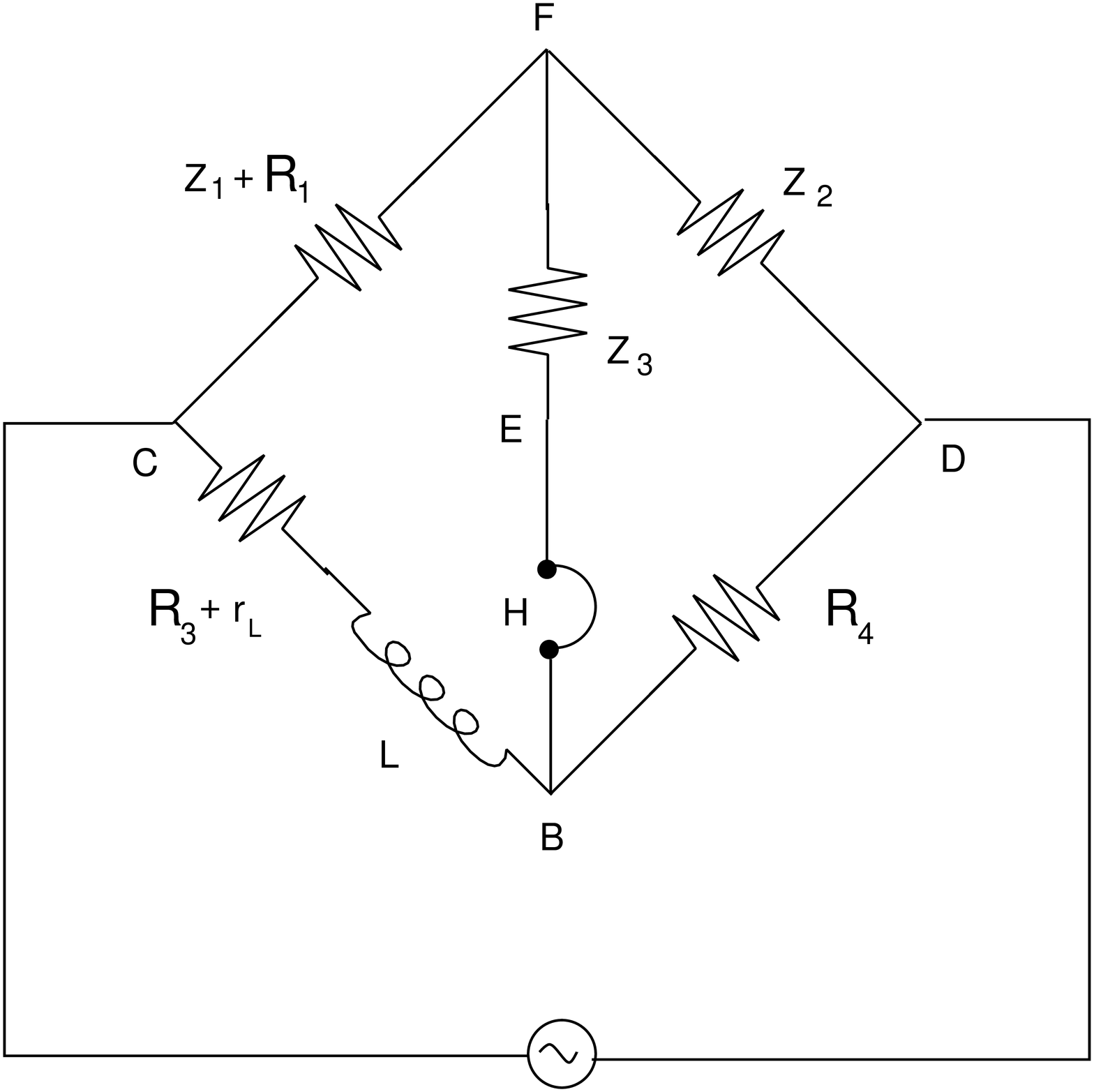}
\caption{The `${\rm \pi}$' shaped circuit between nodes `ADE' is replaced by
an equivalent `T' circuit. On rearranging the circuital elements, the
complex circuit of fig~1 reduces to a apparently simple circuit shown here.
Along with visual simplicity, it renders the computation of potentials
trivial.}  
\label{fig_and_wh} 
\end{center}
\end{figure}
The circuit shown in Fig.~\ref{fig_and} can now be viewed as shown in
Fig.~\ref{fig_and_wh}. The circuit now looks more like the fundamental Wheatstone
bridge, which purely due to its topology looks trivial and overcomes a mental
barrier a student might have due to a complex looking circuit.
\subsection{Balancing Conditions}
 For the bridge to be balanced, the current in arm `EB' should be zero. This
demands the potential ${\rm V_{FB}}$ to be equal to zero. This potential
can be worked out as
\begin{eqnarray}
V_{FB}=V_F-V_B\nonumber
\end{eqnarray}
where, using potential divider expression, we can compute ${\rm V_F}$ and
${\rm V_B}$. They are expressed as
\begin{eqnarray}
V_F &=& \left({Z_2 \over R_1+Z_1+Z_2}\right)V_{DC}\nonumber\\
V_B &=& \left({R_4 \over R_3+r_L+X_L+R_4}\right)V_{DC}\nonumber
\end{eqnarray}
giving the potential ${\rm V_{FB}}$ as
\begin{eqnarray}
V_{FB}= \left({Z_2 \over R_1+Z_1+Z_2}-
{R_4 \over R_3+r_L+X_L+R_4}\right)V_{DC}\label{pot}
\end{eqnarray}
On attaining balance of the bridge
\begin{eqnarray}
{Z_2 \over R_1+Z_1+Z_2}={R_4 \over R_3+r_L+X_L+R_4}\label{pot1}
\end{eqnarray}
which is same as the condition given by the eqn(\ref{wh_bal}) with the four
impedences replaced by the corresponding impedences in the four arms of
Fig.~\ref{fig_and_wh}.

Substituting the expressions of eqn(\ref{z1}), eqn(\ref{z2}) and eqn(\ref{z3})
in eqn(\ref{pot1}), we have
\begin{eqnarray}
{R_2X_c \over R_1(r+R_2+X_c)+rR_2+R_2X_c}=
{R_4 \over R_3+r_L+X_L+R_4}\label{pot2}
\end{eqnarray}
Collecting and simplifying the imaginary terms of the above equation, we get
\begin{eqnarray}
{R_2 \over R_4} &=& {R_1 \over R_3+r_L}\label{dcbal}
\end{eqnarray}
Expression given by eqn(\ref{dcbal}) is the DC balance condition of the
Anderson Bridge. Now collecting the real terms of eqn(\ref{pot2}), we have
\begin{eqnarray}
{L \over C} &=& {R_4 \over R_2}[R_1R_2+r(R_1+R_2)]\label{acbal}
\end{eqnarray}
Eqn~(\ref{acbal}) is the AC balance condition of the bridge.

The unknowns in the two eqns (\ref{dcbal}) and (\ref{acbal}) are $r_L$
 and $L$ respectively. To obtain the D.C. balance of the bridge, the
 circuit is made with `$r$' shorted, capacitance $C$ open, headphone 
replaced by
 a galvanometer and the A.C. source by a battery or D.C. power supply. Now
 $R_3$ is varied until galvanometer shows zero deflection. Eqn.~(\ref{dcbal})
 then gives the value of $r_L$. Now using the complete circuit given in
 Fig.~\ref{fig_and} and  leaving the resistances set ${\rm R_1}$, ${\rm
 R_2}$ and ${\rm R_3}$ undisturbed, AC balance is obtained by varying the 
resistance `$r$'. Eqn.~(\ref{acbal}) can now be used to compute the value of 
self inductance $L$ applied in the circuit. Thus the unknown impedence
 $\sqrt{(L\omega)^2 + r_L^2}$ is computed.
For the simplicity of circuit designing we suggest taking $ R_2=R_4$,
which results in $ R_1=r_L+R_3$ from eqn.~(\ref{pot2}). 

\subsection{Maximum Power Transfer}
For a good audio signal at the head phone, we require that the circuit
transfers a substantial amount of power to the head phone. Since the ear
senses the intensity of sound, the head phone as a measuring device is being
used as a power meter. Good sensitivity hence would be attained as large
power is being transfered to the head phone for both balanced and
unbalanced (AC) condition. Unbalanced AC condition would be represented by
replacing `$r$' (of eqn.~\ref{acbal}) with `$r+dr$'.

\begin{figure}[t!!]
\begin{center}
\includegraphics[width=3in,angle=0]{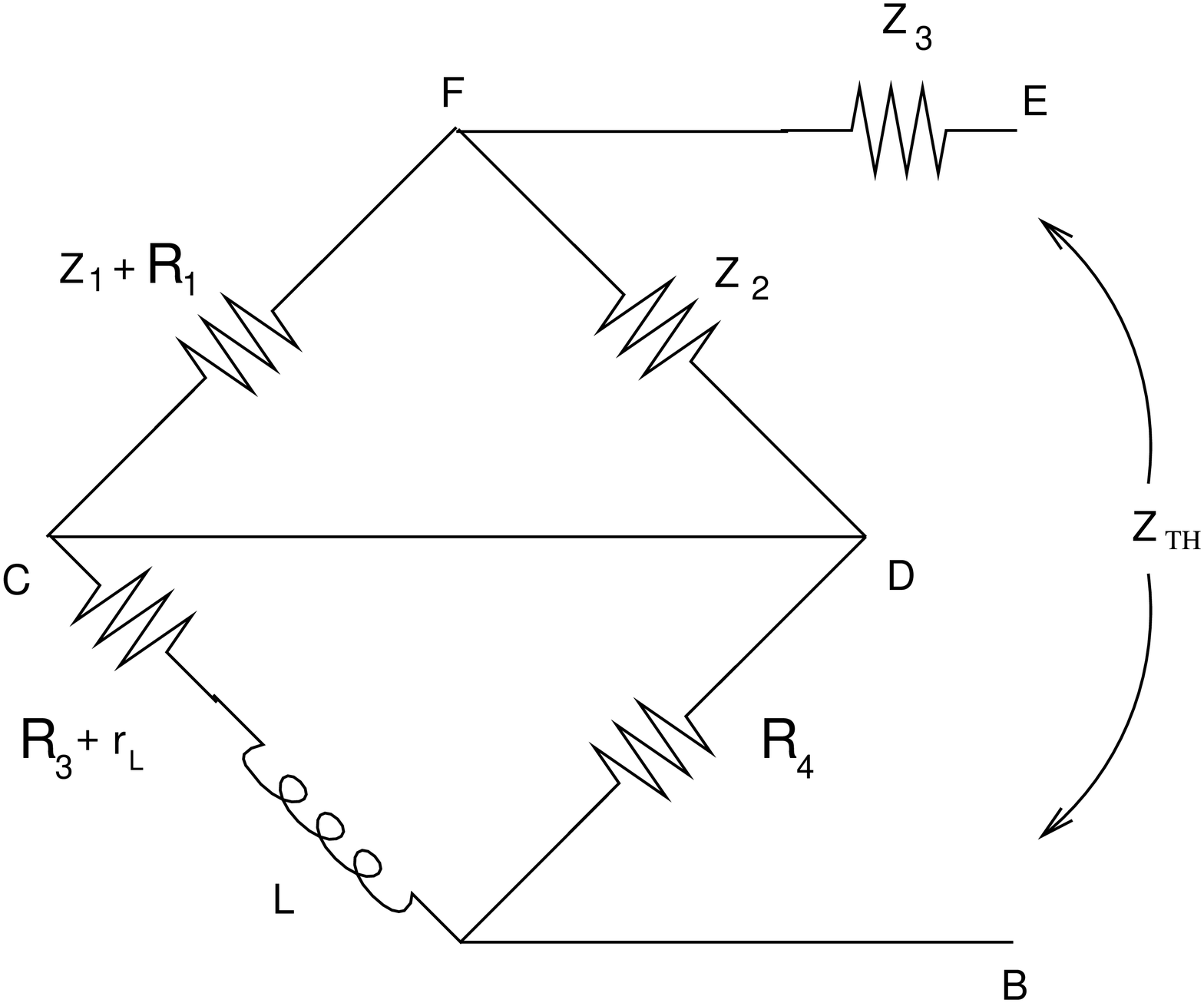}
\caption{The Thevenin impedence is calculated by shorting nodes `C' and `D'
and evaluating the impedence as seen from nodes `EB'.}  
\label{fig_Th_eq} 
\end{center}
\end{figure}

The ``Maximum Power Transfer Theorem'' \cite{edmin} demands that the load
impedance applied to a circuit should be the complex conjugate of the
circuit's Thevenin impedence. Since the head-phone's impedence is inductive in
nature, i.e. it can be represented as `$a+jb$', the theorem demands the
Thevenin impedence should be capacitive in nature, i.e. of the form of
`$a-jb$'. The Thevenin impedence of the Anderson Bridge circuit ($
Z_{\rm {TH}}$) as seen by the head phone is theoretically determined by shorting
the AC source (i.e. we neglect the impedence of the source), giving (see
Fig.~\ref{fig_Th_eq})
\begin{eqnarray}
Z_{TH} &=&\left[ \{(R_1+Z_1)||Z_2 \}+ \{(R_3+r_L+X_L)||R_4 \} + Z_3 \right]
\nonumber\\ 
&=& \left[{(R_1+Z_1)Z_2\over R_1+Z_1+Z_2}+{R_4(R_3+r_L+X_L)\over
R_3+r_L+X_L+R_4}+Z_3\right]\nonumber
\end{eqnarray}
Using the D.C. balance condition given by eqn.~(\ref{dcbal}) and
imposing $R_2 = R_4$, we have
\begin{eqnarray}
Z_{TH} &=& \left[{(R_1+Z_1)Z_2\over R_1+Z_1+Z_2}+{R_2(R_1+X_L)\over 
R_1+X_L+R_2}+Z_3\right]\nonumber
\end{eqnarray}
The best way to simplify the above expression is to select $ X_c \sim
0$. This means the AC audio frequency ($f$) should be kept large and
the capacitor used should have a very large value of `$C$'. From
eqn~(\ref{acbal}), it naturally flows that $ X_L$ has to be
considerably large. Selecting a high audio frequency automatically
achieves this. Further, ${\rm X_c \sim 0}$ might be difficult to
attain given the fact that frequency is to be in audible range and
very high capacitances are usually electrolytic, thus introducing problems
of it's own. Hence it would be sufficient
to select `$C$' and `$f$' such that 
$|X_c| << |X_L|$ and $(rR_1+R_1R_2+rR_2) >> R_2 |X_c|$. Further, let ${\rm
R_1>>R_2}$. The justification and advantage of this assumption
would be evident later. With these conditions 
imposed, the Thevenin equivalent
impedence reduces to
\begin{eqnarray}
Z_{TH} &=& \left[R_2+{(R_2 +r)X_c\over (r+R_2+ X_c)}\right]\label{thev}
\end{eqnarray}
Thus, the basic constraints imposed on the circuit designing has made
the Bridge's Thevenin equivalent impedence capacitive in nature,
(eqn.~\ref{thev} is of the form a-jb).  
\begin{eqnarray}
Z_{TH} &=& \left[R_2+{(r+R_2)|X_c|^2 \over
(r+R_2)^2+|X_c|^2}\right]+\left[{(r+R_2)^2 \over
(r+R_2)^2+|X_c|^2}\right]X_c\label{thev1}
\end{eqnarray}
As stated earlier, for the maximum
power to be transferred, the Bridge's Thevenin equivalent impedence
should be capacitive in nature since the head-phone's impedence is
inductive in nature. Also, head phones are usually low impedence
devices and hence ${\rm X_c}$ and ${\rm R_2}$ should be kept as small
as possible. While selection of small ${\rm X_c}$ was already imposed,
an additional requirement is demanded, i.e. ${\rm R_2}$ should also be of
small value.

\subsection{Sensitivity}
To appreciate the approach of the AC balance point, the sensitivity of the
circuit should be good. Here we define the bridge's sensitivity as (dP/dr),
i.e. the power developed across the head-phone should show large change for a 
small change in `$r$'. Upon the application of the {\sl Maximum Power Transfer 
theorem}, the expression of current can be written as
\begin{eqnarray}
i &=& \frac{V_{TH}}{2 Re(Z_{TH})}\nonumber
\end{eqnarray}
where ${\rm V_{TH}}$ is the Thevenin equivalent voltage. If the
Thevenin impedence doesn't match the impedence of the headphone exactly, 
the current expression for a
circuit designed along the lines discussed above 
reduces to (the Thevenin voltage would be of the same form as that given by 
eqn.~\ref{pot}. Notice ${\rm V_{DC}}$ has been replaced by ${\rm V_{ac}}$, 
the applied A.C. voltage) 
\begin{eqnarray}
i=\left[{1 \over \left({R_H + {r |X_c|^2 \over r^2+|X_c|^2}}\right)+
j\left({Z_H - {r^2 |X_c| \over
r^2+|X_c|^2}}\right)}\right]V_{ac}'\label{potthev1}
\end{eqnarray}
where ${\rm R_H}$ and ${\rm Z_H}$ are the real and imaginary part of the
impedence associated with the head phone being used. The term ${\rm
V_{ac}'}$ is given as 
\begin{eqnarray}
V_{ac}'=\left({Z_2 \over R_1+Z_1+Z_2}-{R_4 \over R_3+r_L+X_L+R_4}\right)V_{ac}
\label{potthev}
\end{eqnarray}
Using the constraints imposed and eqn~(\ref{thev1}), this expression reduces to 
\begin{eqnarray}
V_{ac}'=R_2\left[{X_c \over R_1(r+R_2+X_c)}-{1 \over R_1+X_L}\right]V_{ac}
\nonumber
\end{eqnarray}
Hence, the power dissipated across the head phone is given as
\begin{eqnarray}
P &=& ii^*R_H\nonumber\\
P &\approx& \left[{R_H \over \left({R_H + {r |X_c|^2 \over
r^2+|X_c|^2}}\right)^2+
\left({Z_H - {r^2 |X_c| \over
r^2+|X_c|^2}}\right)^2}\right]V'^2_{ac}\nonumber
\end{eqnarray}
Differentiating w.r.t. `$r$' (for simplifying the expression we can assume
${\rm R_H}$ and ${\rm Z_H \sim 0}$ which is generally true since head phones
are low impedence devices. Further approximation can be made that ${\rm
V_{ac}'}$ is very small and hence is a shallow function of `r' due to the
small values of ${\rm R_2}$ and ${\rm X_c}$ in the numerator and large ${\rm
R_1}$ and ${\rm X_L}$ in the denominator.),
\begin{eqnarray}
{1\over R_HV_{ac}'^2}\Biggr\vert\left({dP \over dr}\right)\Biggr\vert
&\approx& {2 \over r^3}\label{new} 
\end{eqnarray}
The above expression demands `r' to be of small value for good sensitivity
of the bridge. However, till now we have been silent on the nature of `r'.
For this we revisit eq~(\ref{acbal}), to understand the effects of our 
considerations
on the nature of `r'. For a bridge designed along our suggestions, the AC
balance condition (eq~\ref{acbal}) gives
\begin{eqnarray}
r \approx {L \over R_1C} \approx X_c\left({X_L \over R_1}\right)\label{acbalr}
\end{eqnarray}
The selection of high frequency, large capacitance with large ${\rm R_1}$
makes sure that `r', the resistance used to AC balance
the bridge, is proportional to the impedence offered by the capacitor. As
the case is, in our design AC balance should be achieved with small `r'
which would result in good sensitivity. The interpretations do not vary even
if the approximations listed above eqn~{\ref{new} are not made.

 Another design consideration necessary is to make $ R_1$ large. 
We summarize here the circuitial conditions one must 
maintain to get good
sensitivity for the Anderson Bridge
\begin{itemize}
\item[1.] $(R_1=R_3) >> (R_2=R_4)$ with $ R_2$ being very small
\item[2.] Select ${\rm X_L >> X_c}$.
\item[3.] Audio frequency to be large
\end{itemize}

We put our analysis to test by designing an Anderson Bridge subject to
the conditions listed above and determined the self inductance of a
given coil. We report some select observations in Table 4 to highlight
our findings. The headphone used as a current detecting instrument had
a resistance $R_H$ of about $ 100 \Omega$ and inductance $L_H$ of
about $22 mH$. The first reading in Table 4 shows our best result,
where we ensured that all the conditions required for proper design
alongwith Maximum Power Transfer Theorem are satisfied. The ability in
determining the balance point, in terms of lowest audio internsity,
was also remarkable. For increase of ${\rm \pm 1 \Omega}$ in the first
reading we could detect increase in sound intensity. For the second
reading we took a capacitor ten times smaller than the capacitor used
in the previous set of observations.  This ofcourse results in a ten
fold increase in the value of $ X_c$ and `r'.  The values of other impedences
selected here also adhered to the listed conditions. However, the
inaccuracy in the value of inductance determined is evident. Even the
ability to determine the balance point was compromised in the second
design. Similar, inaccuracy is also evident when the audio frequency
is decreased to $1 kHz$ in the third design reducing the value of
$X_L$ and increasing that of $X_c$.  
\begin{table}[h]
\begin{center}
\caption{\sl Table lists the experimentally determined value of
inductance (L) in bridge with coil L=130mH and circuital elements selected
as per the analysis described above.}
\vskip 0.5cm
\begin{tabular}{|c|c| c| c| c| c| c| c| }
\hline
S.No. & f (KHz) & ${\rm R_1\, (\Omega)}$ & ${\rm R_2\, (\Omega)}$
& ${\rm R_3\, (\Omega)}$ & C (${\rm \mu F}$) & 
r (${\rm \Omega}$) & L (mH)  \\ \hline
1. & 3 & 4700 & 100 & 4770 & 0.10 & 186-188 & 136.7  \\ \hline
2. & 3 & 4700 & 100 & 4770 & 0.01 & 5570-5720 & 277.5 \\ \hline
3. & 1 & 4700 & 100 & 4770 & 0.10 & 293-313 & 192.6 \\ \hline
\end{tabular}
\label {tab}
\end{center}
\end{table}

The design constraints discussed here were not discussed with the
experimentalist who reported results of Tables 1, 2 and 3. On examination of 
the the data, we find that only the first reading of Table 2 and Table 3 
satisfy stated conditions of circuit design. Not only are these the best
results in terms of returned value of inductance but also in terms of
ability to resolve the minimum sound. 

\section*{Conclusion}
Based on our derivations using various theorems of {\sl Network
Analysis}, we have analysed how to design an Anderson Bridge with good
sensitivity. Varous considerations show that for a better sensitivity,
one should work at high frequencies and use a low $R_2$ and high
capacitance (making $X_L >> X_c$). We hope that the present treatment 
allows the
circuit not just to serve as a method of determining self inductance
but also becomes an important pedagogical tool for circuit analysis.
\section*{Acknowledgement}
The authors would like to thank their colleagues Dr. Deepak Chandra
and Dr. Ravi S. Bhattacharjee for fruitful discussions. We also
express our sincere gratidute to Mr. Vikas Nautiyal, whose
experimental test results are listed in Table 1, 2 and 3.  The
assistance of the technical staff of Department of Physics and
Electronics of S.G.T.B. Khalsa College is also acknowledged.
 
%
\end{document}